\documentclass[
  3p
]{elsarticle}

\usepackage{graphicx}
\usepackage{dcolumn}
\usepackage{bm}
\usepackage{hyperref}
\usepackage{url}
\usepackage{comment}
\usepackage{xparse}
\usepackage{amsmath}

\newcommand{\eref}[1]{Eq.~(\ref{#1})}
\newcommand{\Eref}[1]{Equation~(\ref{#1})}
\newcommand{\fref}[1]{Fig.~\ref{#1}}
\newcommand{\Fref}[1]{Figure~\ref{#1}}
\newcommand{\tref}[1]{Table~\ref{#1}}

\newcommand{\appropto}{\mathrel{\vcenter{
  \offinterlineskip\halign{\hfil$##$\cr
    \propto\cr\noalign{\kern2pt}\sim\cr\noalign{\kern-2pt}}}}}

\begin{document}
\newcommand{\ManuscriptTitle}{
    Plasma-Parameter Dependence of Ro-Vibrational Temperatures \\
    for $\mathrm{H}_2$ in LHD Divertor
}

\title{\ManuscriptTitle}


\author[kyoto,ornl]{Keisuke Fujii}
\ead{fujiik@ornl.gov}
\author[kyoto]{Tsubasa Oshioka}
\author[kyoto]{Atsushi Niihama}
\author[kyoto]{Kuzmin Arseniy}
\author[kyoto]{Taiichi Shikama}
\author[nifs]{Masahiro Kobayashi}
\author[kyoto]{Masahiro Hasuo}
\author{the LHD Experiment Group}
\address[ornl]{Oak Ridge National Laboratory, Oak Ridge, TN 37831-6169, United States of America}
\address[kyoto]{%
Department of Mechanical Engineering and Science,
Graduate School of Engineering, Kyoto University
Kyoto 615-8540, Japan}
\address[nifs]{%
National Institute for Fusion Science, Toki, Gifu, 5909-5292, Japan}

\date{\today}

\begin{abstract}    
We analyzed a thousand visible spectra of Fulcher-$\alpha$ band measured for divertor plasmas in Large Helical Device.
With a coronal model and Baysian inference, the population distribution of hydrogen molecule in the electronical ground state were estimated.
The non-thermal population distribution was recovered with a two-temperature model, which has two sets of rotational and vibrational temperatures, as well as their mixture coefficient.
The lower rotational temperature significantly changes according to the plasma parameters. 
Its nearly linear dependence on the electron density was found, which is consistent with previous works.
The lower vibrational temperature also shows a small density dependence, as reported by a previous work. 
On the other hand, the higher rotational and vibrational temperatures as well as the mixture coefficient only show slight changes over the broad range of plasma parameters. 
These population parameters show a significant correlation; with higher electron density, all the temperatures and the fraction of the higher-temperature component increase simultaneously. 
This suggests that the electron-impact plays an important role to determine the population distribution.
\end{abstract}

\maketitle

\section{Introduction}
Hydrogen molecules and their isotopes play a key role on chemical reactions in the divertor of magnetic fusion devices.
The rates of many molecular relevant processes, such as the formation of negative ions~\cite{bacal_pressure_1981,gorse_dynamics_1992}, dissociation and ionization~\cite{Sawada1993}, and dissociative attachment~\cite{capitelli_open_2005,may_non-linear_nodate,krasheninnikov_plasma-neutral_1997}, are known to be significantly influenced by the rotational and vibrational population distribution in the electronic ground state of hydrogen molecules.
Rotational and vibrational populations in the electronic ground state ($X$ state) have been estimated for variety of plasmas,
based on the direct measurement with vaccuum ultraviolet lasers~\cite{pealat_rovibrational_1985,stutzin__1990,vankan_high_2004,gabriel_formation_2008} or the combination of visible emission observations and population-kinetics models (e.g., a corona model and collisional-radiative model)~\cite{xiao_rovibrational_2004,fantz_spectroscopypowerful_2006,Briefi2017}.
In many cases, non-thermal population distribution has been observed, where the vibrational temperature is significantly higher than the rotational temperature~\cite{xiao_rovibrational_2004} and highly excited rotational states are excessively populated than expected from a Boltzmann distribution~\cite{vankan_high_2004,gabriel_formation_2008, Briefi2017}.
A Monte-Carlo simulation has also predicted the nonthermal population distribution of hydrogen molecules~\cite{Sawada2020-er}.


The rotational temperature, which is estimated from the population at the low-rotational-quantum-number states, has been thought close to the wall temperatures~\cite{Stutzin1989-oj,Watts2001-up}, because the energy gap among the rotational states is in the similar scale to the room temperature.
Addition to the wall effect, cascade from upper electronic levels may also change the rotational  temperature. 
This effect has been attributed to account the electron-density dependence of the rotational temperature found in the spectroscopic observation of magnetic plasma confinement devices~\cite{brezinsek_molecular_2002,unterberg_local_2005,Hollmann2006-zy}.

A surface effect has been considered as the origin of the higher vibrational temperature.
The surface assisted recombination of two hydrogen atoms results in the highly excited molecules~\cite{capitelli_open_2005}.
Although, a plasma-condition dependence on the vibrational temperatures have been studied~\cite{fantz_spectroscopic_1998}, the connection to the elementary processes are still unknown.

Despite the understandings of such microscopic processes, the full understandings of population kinetics are still lacking.
Experimental evaluations of the population kinetics in variety of plasmas may help us to reveal the macroscopic behavior.
%
In this work, we study the emission spectra of hydrogen Fulcher-$\alpha$ band, which are routinely obtained from the divertor region of LHD.
The spectra were observed with an echelle spectrometer that has both high wavelength resolution $\approx$~0.1~nm and wide wavelength bandwidth 400--760 nm. 
We analyzed a thousand of spectra observed for a hundred of LHD discharge experiments, covering wide range of plasma parameters.
By assuming a two temperature model and utilizing the Bayesian inference framework, we robustly estimated the population distribution of the electronic ground state of hydrogen molecules for these experiments.
We found that the lower rotational temperature shows a significant dependence on the plasma parameters, as reported in previous works~\cite{brezinsek_molecular_2002,unterberg_local_2005}. 
On the other hand, the other parameters show only a little change over the parameter range we studied.
Furthermore, significant correlation among these temperatures is found, suggesting an existence of a simple mechanism generating the non-thermal population distribution. 
For example, Sawada et al have pointed out that with the electron-impact excitation similar nonthermal population distribution can be formed in plasmas~\cite{sawada_rovibrationally_2016}.
Our result is qualitatively consistent with their simulation.

The rest of this paper is organized as follows.
In section~\ref{sec:experiment}, we briefly present our experimental setup, although the same setup was used in our preceding paper~\cite{ishihara}.
In section~\ref{sec:analysis}, the analysis method and our assumption are described. 
The result for typical spectra are also shown in this section.
In section~\ref{sec:correlation}, we present our full results for the thousands frames.
The detailed correlation among the estimated parameters are presented.

\section{Experimental Setup\label{sec:experiment}}

\subsection{Large Helical Device}
\begin{figure}[hbt]
  \includegraphics[width=8cm]{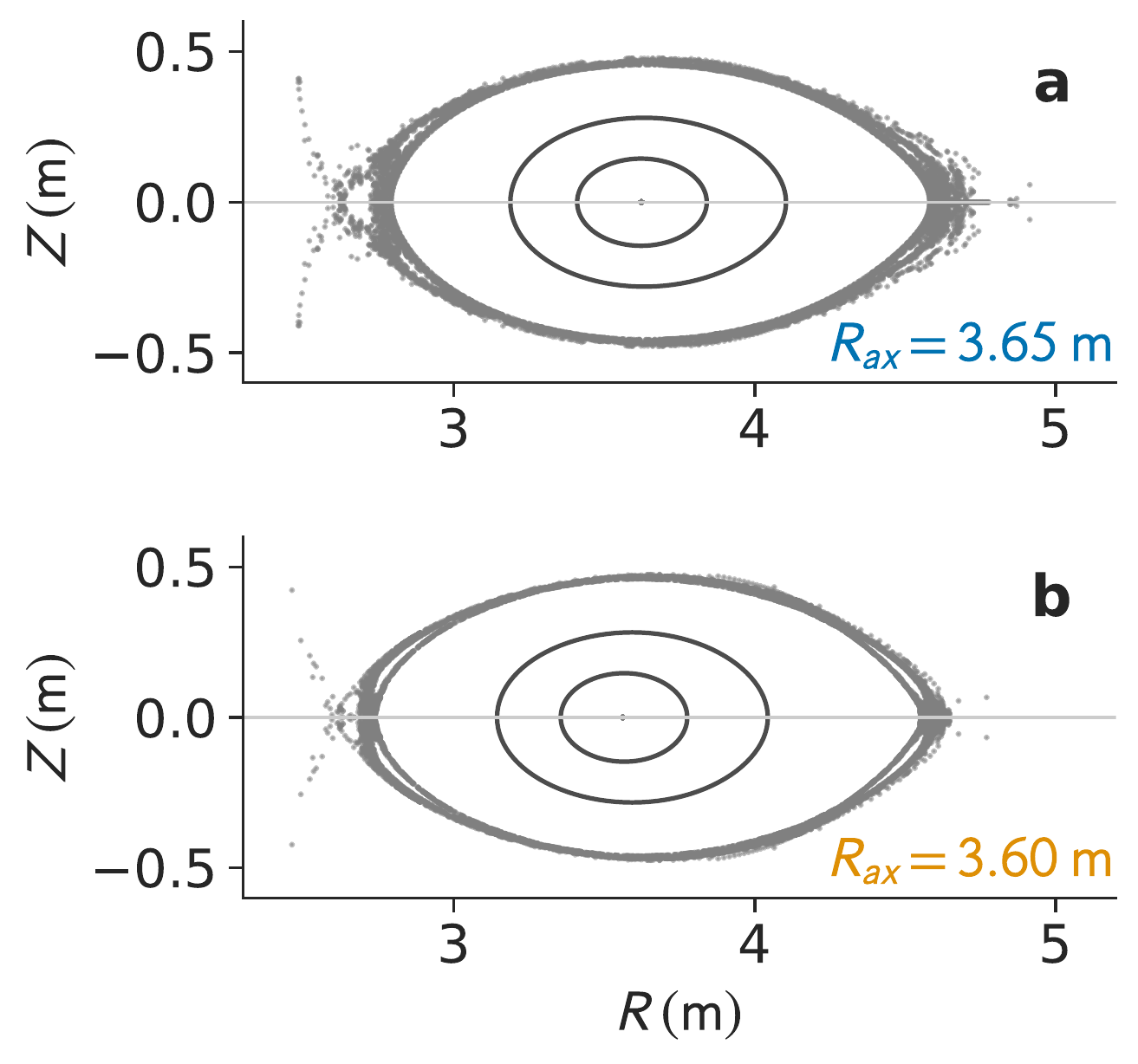}
  \caption{%
  Poloidal crosssections of LHD plasmas for (a) the configuration $R_{ax} = 3.65$ m and (b) $R_{ax} = 3.60$ m.
  Some closed magnetic flux surfaces are shown by solid curves while the stochastic layer encircling the last closed flux surface is shown by gray dots.
  The stochastic layer also provides helical divertor structure, which connects the main plasma and divertor located in the inner and outer boards.
  }
  \label{fig:poloidal}
\end{figure}

LHD is a helical-type magnetic-plasma-confinement machine.
With a pair of helical coils, plasmas with the major radius $R = $ 3.5--4.0 m and the minor radius $\approx 0.6$ m are generated.
This helical coil also makes helical divertor structures.
A poloidal cross section of the LHD plasma is shown in \fref{fig:poloidal}.
In the figure, the closed magnetic flux surfaces are indicated by gray curves.
The cross points of open magnetic field lines are shown by gray dots.
The intrinsic helical divertor structure and the stochastic layers can be seen.
Inside the last closed flux surface (LCFS), ions and electrons are confined.
Once these charged particles diffuse out the LCFS, they are guided by an open magnetic field line to divertor plates through the divertor leg structure.

Depending on the coil current, LHD can take several magnetic configurations.
The magnetic axis position $R_{ax}$ is an important parameter representing a configuration.
With the inner shifted configuration $R_{ax} = 3.60$ m (\fref{fig:poloidal}~(b)), the LHD plasma has thinner ergodic layer and a less helical ripple than those with $R_{ax} = 3.65$ m (\fref{fig:poloidal}~(a)).

The plasma is heated by neutral-beam-injections (NBIs) and electron cyclotron heating (ECH). 
LHD equips several gas injection systems.
By varying the gas pressure, heating power, and the magnetic configuration, variety of plasmas spanning broad range of parameter space can be generated.

\begin{figure}[ht]
  \includegraphics[width=8cm]{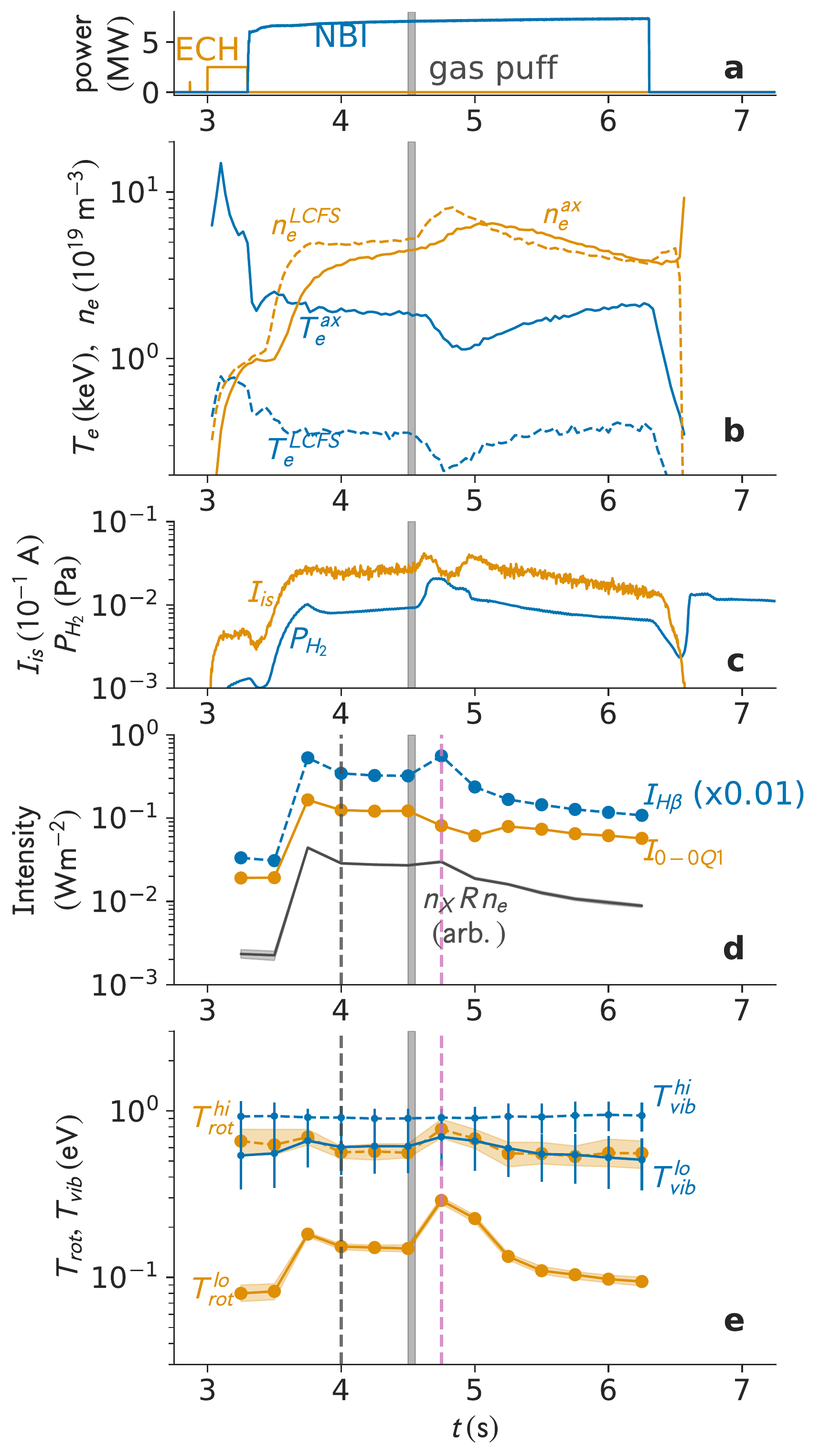}
  \caption{%
  Temporal evolution of a typical experiment (shotnumber \#142897). 
  (a) Heating power by neutral beam injection (NBI) and electron cyclotron heating (ECH).
  (b) Temporal evolutions of $T_e$ and $n_e$ on the plasma axis ($T_e^{ax}$ and $n_e^{ax}$, respectively) and last closed flux surface ($T_e^{LCFS}$ and $n_e^{LCFS}$, respectively).
  Because of the gas injection at $t$ = 4.5 s (indicated by the thick vertical bar), $n_e$ increases and $T_e$ decreases.
  (c) Temporal evolutions of the ion saturation current measured at the divertor plate ($I_{is}$) and neutral gas pressure measured near the divertor region ($P_{H2}$).
  (d) Measured emission intensities of hydrogen atomic Balmer-$\beta$ ($I_{H_\beta}$) and the Fulcher-$\alpha$ $(0-0) Q1$ transition.
  The estimated result of the excitation flux ($n_X\,R\,n_e$, see the text for its definition) is also shown.
  (e) Estimated rotational and vibrational temperatures of the $X$-state of hydrogen molecule.
  Estimation uncertainties are also shown either by the error bars and width of the curves.
  }
  \label{fig:shotsummary}
\end{figure}

An example of the temporal sequence of a discharge experiment is shown in \fref{fig:shotsummary}.
This plasma is initiated by ECH and sustained by NBI (\fref{fig:shotsummary}~(a)).
In \fref{fig:shotsummary}~(b), the temporal evolutions of the electron temperature ($T_e$) and density ($n_e$) measured by Thomson scattering method are shown.
In this experiment, helium gas is injected at $t$ = 4.5 s, resulting in an increase in $n_e$ and decrease in $T_e$, followed by their gradual recoveries.

\begin{figure}[ht]
  \includegraphics[width=8cm]{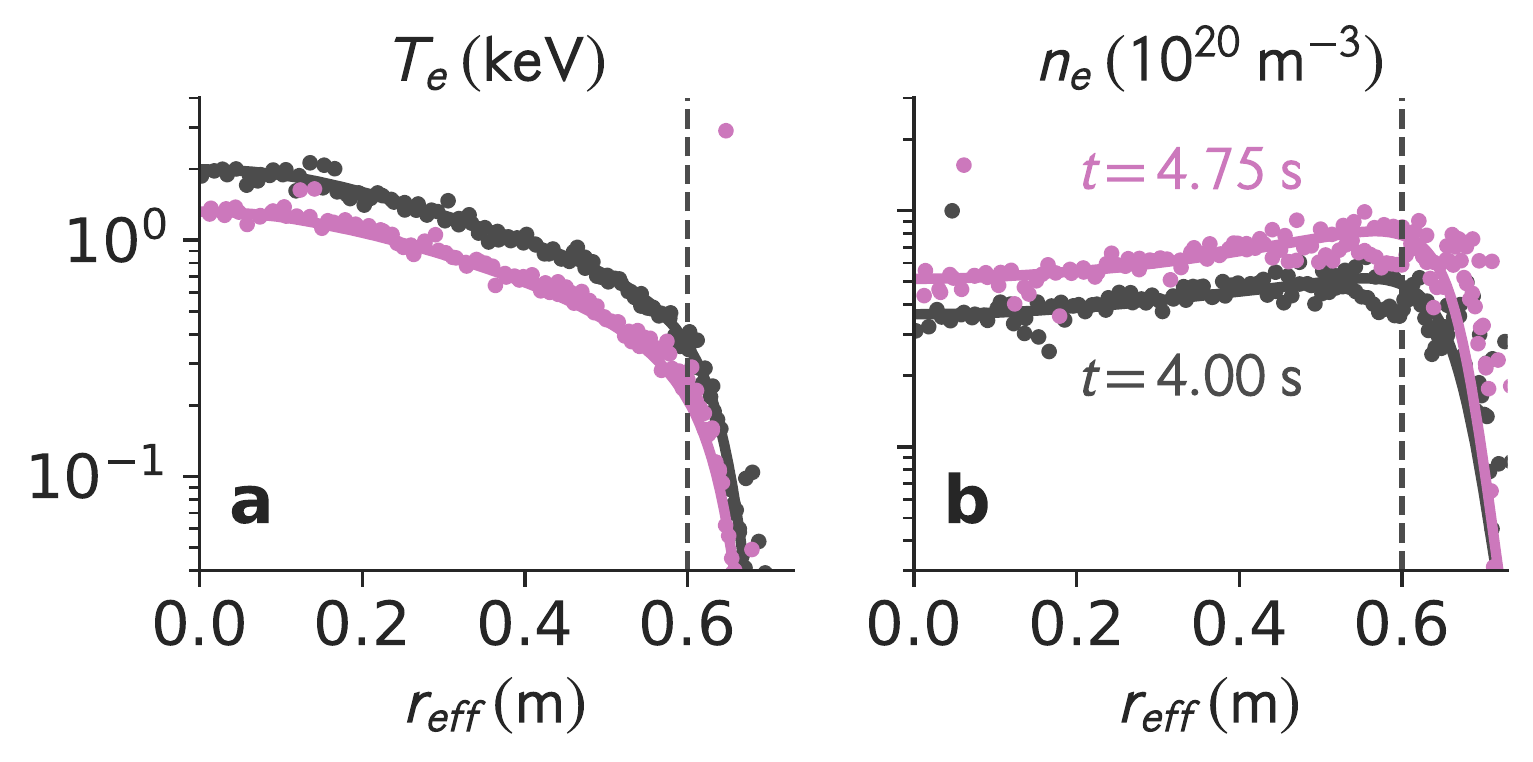}
  \caption{%
  Radial distributions of (a) $T_e$ and (b) $n_e$ measured for \#142897 at $t=$ 4.00 s and 4.75 s with Thomson scattering method.
  Rough position of the LCFS is shown by vertical dashed lines.
  }
  \label{fig:thomson}
\end{figure}

This Thomson scattering system measures the radial distributions of $T_e$ and $n_e$ with a spatial resolution of $\approx 10$~mm and temporal resolution of $\approx 30$~ms.
The spatial distributions of $T_e$ and $n_e$ measured at $t$ =~4.00 and 4.75~s are shown in \fref{fig:thomson}.
Although it is difficult to precisely determine the position of the LCFS due to the stochastic nature of the magnetic field lines at the edge region, typically this is located at $r_{eff} \approx 0.6$ m, where $r_{eff}$ is the effective minor radius.
$T_e^{ax}$ and $n_e^{ax}$ shown in \fref{fig:shotsummary}~(b) are the values of $T_e$ and $n_e$ at the magnetic axis $r_{eff}$ = 0 m, while $T_e^{LCFS}$ and $n_e^{LCFS}$ indicate these values at $r_{eff}$ = 0.6 m.

Many other diagnostics have been installed in LHD. 
In this work, we consider some of old-established diagnostics, namely 
the ion saturation current onto divertor plates $I_{is}$ measured by Langmuir probes, and the gas pressure measured in the divertor region $P_{H_2}$ by a fast-ionization-guage, in addition to the $n_e$ and $T_e$ values measured by the Thomson scattering method.
The temporal evolutions of the $I_{is}$ and $P_{H_2}$ are shown in ~\fref{fig:shotsummary}~(c).
Because of the gas injection, the gas pressure increases.
The ion saturation current shows a non-monotonic behavior, indicating the interplay between the increase in $n_i$ and decrease in $T_i$ in the diverotr region.

\subsection{Visible spectroscopy}

We observed the visible emission from the divertor region of the LHD.
The experimental setup is the same with that shown in Ref.\cite{ishihara}, where the emission at the inner divertor region was collected by an optical lens, focused on an optical fiber edge, transferred to the entrance slit of an echelle spectrometer which have been developed by our group~\cite{Tanaka2020-gg,Hasuo2012-dv}.
This spectrometer measures the spectrum in the wavelength range of 400--780 nm with the wavelength resolution of $\approx 0.1$ nm simultaneously.
33 ms exposure time and 4 Hz frame rate are used for all the results shown in this work.

\begin{figure*}[ht]
  \includegraphics[width=18cm]{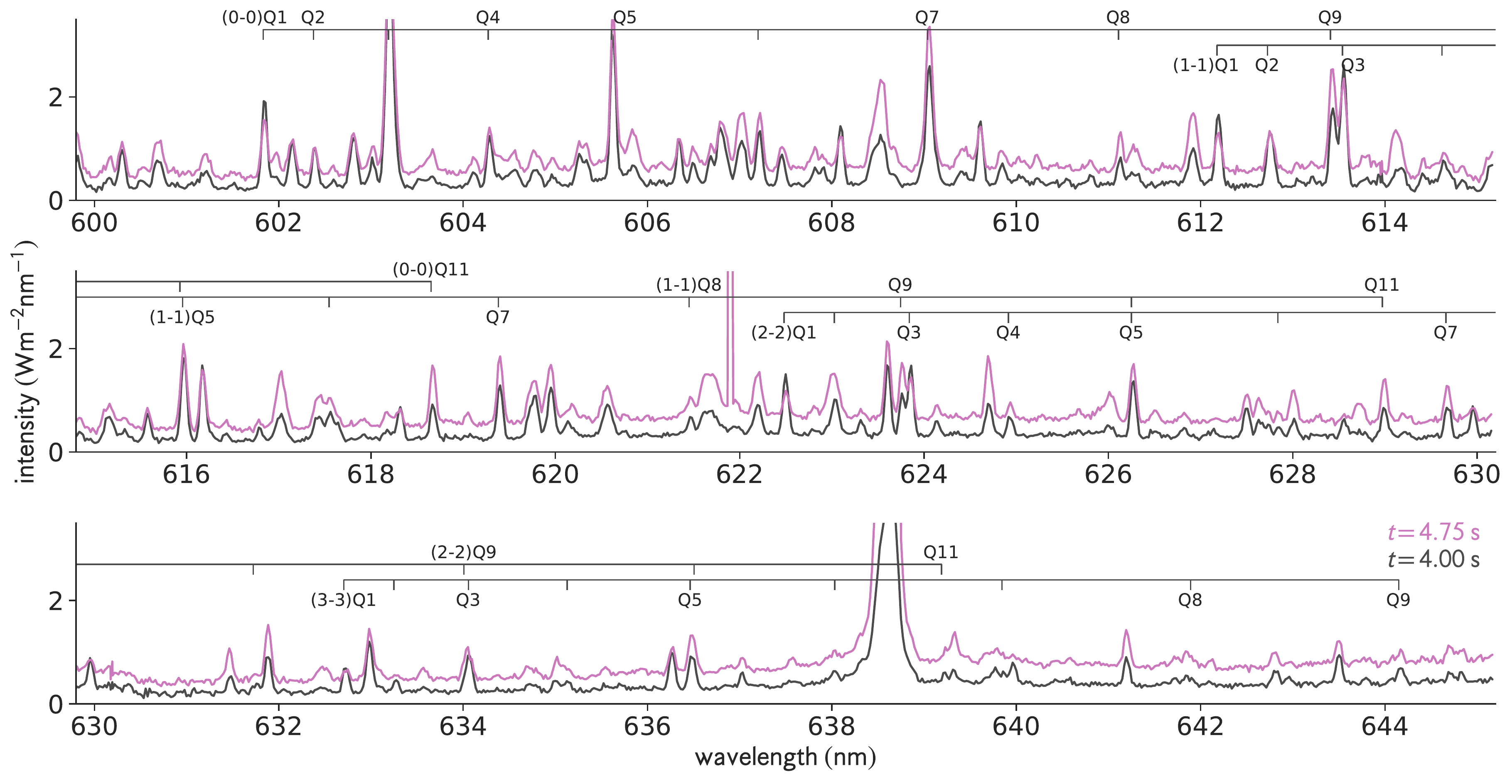}
  \caption{%
    The emission spectra observed at $t$ = 4.0 and 4.75 s with the echelle spectrometer.
    The central wavelengths of Fulcher-$\alpha$ band are shown by the vertical bars.
    The transitions for the emission lines used in the analysis are also indicated.
  }
  \label{fig:spectra}
\end{figure*}

\Fref{fig:spectra} shows the visible spectra measured by this spectrometer at $t$ = 4.00 and 4.75 s for the experiment shown in \fref{fig:shotsummary}.
The vertical bars in the figure show the central wavelengths of the Q branches of the hydrogen Fulcher-$\alpha$ band.
The Fulcher-$\alpha$ band is the emission lines from $d ^3\Pi_u^-$ state to $a ^3 \Sigma_g^+$ state.
The transition notations $(v'-v'') QN'$ are shown in the figure, where $v'$ and $v''$ indicate the vibrational quantum numbers of the upper and lower states, respectively, while $N'$ indicates the rotational quantum number of the upper state.
As it is in the $Q$ branch, $N' = N''$.
For example, $(0-0) Q1$ indicates $v'=0, v''=0$, and $N'$ = $N''$ = 1.

After the gas injection, the intensities of these lines change.
The intensity of the $(0-0) Q1$ line decreases while that of the $(0-0) Q11$ line increases.
We estimate the emission intensities by fitting the spectra by a Gaussian function.
The temporal evolution of the $(0-0) Q1$ line intensity is shown in \fref{fig:shotsummary}~(d).
In this panel, we also plot the intensity evolution of the hydrogen atom Balmer-$\beta$ line.
Balmer-$\beta$ line intensity increases according to the gas injection while the $(0-0) Q1$ line intensity decreases.

We estimate the population of the $d$ states from the intensity $I_{(v'-v'')QN'}$ based on the following relation
\begin{align}
  I_{(v'-v'')QN'} = h\nu^{d v' N'}_{a v'' N''}\; A^{d v' N'}_{a v'' N''}\; n_{d v' N'},
\end{align}
where $\nu^{d v' N'}_{a v'' N''}$ is the photon energy of the $(d, v', N') \rightarrow (a, v'', N'')$ transition and $n_{d v' N'}$ is the line-integrated population density of the upper state.
$A^{d v' N'}_{a v'' N''}$ is the Einstein coefficient for the transition $d v' N'\rightarrow a v'' N'' $,which are computed by~\cite{Surrey2003-st}
\begin{align}
    A^{d v' N'}_{a v'' N''} = 
    \frac{16 \pi^3}{3h^4\epsilon_0 c^3}
    (h\nu^{d v' N'}_{a v'' N''})^3
    \,
    \overline{R_e}^2
    \,
    q^{dv'}_{av''}
    \,
    \frac{S_{N'N''}}{2N' + 1}
    \label{eq:fc}
\end{align}
with the vecuum permittivity $\epsilon_0$, the planck constant $h$, the light speed $c$, the dipole transition moment $\overline{R_e}$ between $a$ and $d$ states, the Franck-Condon factor between the upper and lower vibrational states $q^{dv'}_{av''}$, and the H\"onl-London factor for the $Q$ branch $S_{N'N''}=(2N'+1) / 2$.
The values of $q^{dv'}_{av''}$ are taken from Ref.~\cite{fantz_franckcondon_2006}.

\begin{figure}[ht]
  \includegraphics[width=8cm]{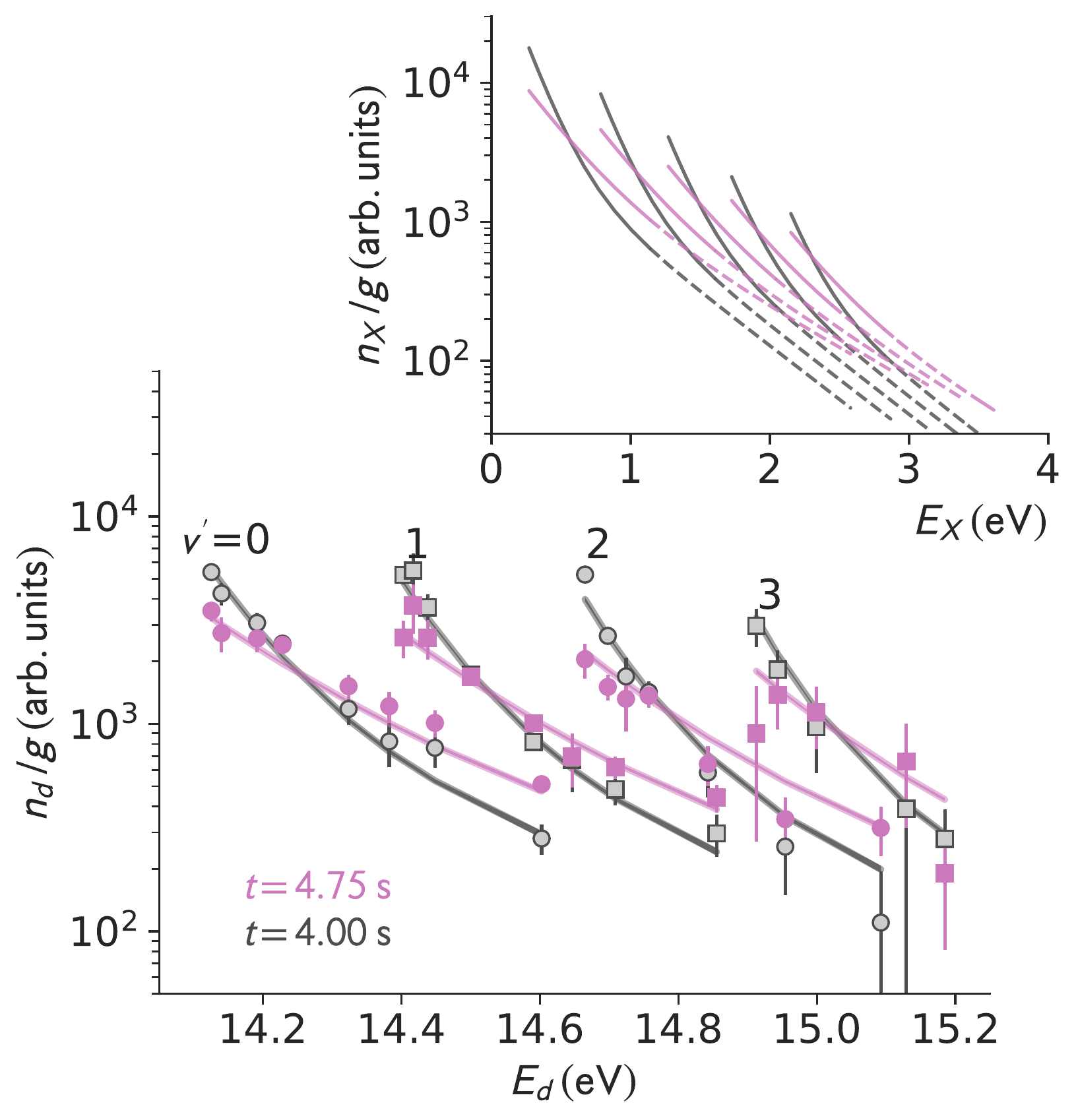}
  \caption{%
  The population of $d$-state observed at $t$ = 4.0 and 4.75 s.
  The solid curves show the best fit by the coronal model (\eref{eq:coronal_model}) with the two-temperature Boltzmann distribution for $X$ state (\eref{eq:two_temperature}).
  The estimated $X$ state populations are shown in the inset.
  }
  \label{fig:population}
\end{figure}

\Fref{fig:population} shows the population of the upper state of the Fulcher-$\alpha$ band, normalized by the statistical weight of the upper state $g_{N'} = (2N'+1)g_{as}$ with the nucleus statistical weight $g_{as} = 2 - (-1)^{N'}$.
The rotational population for each $v'$ state shows the decreasing trend as a function of the excited energy.
The slope in the vertical logarithmic scale is steeper at the lower density plasma ($t$ = 4.00 s) and more flat at the higher density plasma ($t$ = 4.75 s).

\section{Estimation of the Population Distribution of the $X$ states\label{sec:analysis}}
Only the upper-state population can be directly obtained from the observed emission intensities.
In order to estimate the population distribution in the $X$ state, we utilize the coronal model, which has been widely used in the literature~\cite{xiao_rovibrational_2004,fantz_spectroscopypowerful_2006,Briefi2017,fantz_spectroscopic_1998}.
Note that the coronal model holds only when the electron density is sufficiently small. 
In high-density plasmas, the population influx from excited states becomes important, however, the validity criteria is not yet established. 
Thus, in this work we ignore the effect and assume that the coronal model is valid in our parameter range.

\subsection{Coronal model}
With the assumption of the coronal equilibrium, the population distribution of the $d$ state is determined by the population influx from the $X$ state ($\Gamma_{X v N}^{d v' N'}$) and the outflux to the $a$ state ($\Gamma_{d v' N'}^{a v'' N''}$).
$\Gamma_{X v N}^{d v' N'}$ is computed with the Born-Oppenheimer approximation,
\begin{align}
    \Gamma_{X v N}^{d v' N'} \approx q^{Xv}_{dv'} \delta(N-N') \,
    n_{X v N} \,
    R(T_e)\,
    n_e.
\end{align}    
where $q^{Xv}_{dv'}$ is the Franck-Condon factor between $X$ and $d$ states, and $R(T_e)$ is the excitation rate coefficient from $X$ to $d$ states.
The population outflux $\Gamma_{d v' N'}^{a v'' N''}$ is
\begin{align}
    \Gamma_{d v' N'}^{a v'' N''} \approx A^{d v' N'}_{a v'' N''} \delta(N'-N'')
      \; n_{d v' N'}.
\end{align}
From the steady state condition 
$\sum_{v,N}\Gamma_{X v N}^{d v' N'} = \sum_{v''}\Gamma_{d v' N'}^{av''N''}$, 
the population of the $d$ state can be written by
\begin{align}
  n_{d v' N'} = \frac{
      \sum_{v, N} \Gamma_{X v N}^{d v' N'}
    }{
      \sum_{v''} A^{d v' N'}_{a v'' N'}
    }.
  \label{eq:coronal_model}
\end{align}

\subsection{Ro-Vibrational Temperature Estimation of the $X$ state}

Because \eref{eq:coronal_model} involves more number of unknown parameters ($n_{X v N}$) than the measured values of $n_{d v' N'}$, an appropriate parameterization is necessary to estimate the distribution of $n_{X v N}$~\cite{fantz_spectroscopic_1998}.
Although Boltzmann's distribution has been assumed for $n_{X v N}$ in several works~\cite{xiao_rovibrational_2004,fantz_spectroscopypowerful_2006,Briefi2017,fantz_spectroscopic_1998}, it has been also known that the distribution deviates from the Boltzmann distribution, particularly in the high rotational-quantum-number states.
In fact, our observation of the population distribution shown in \fref{fig:population} also presents a deviation from the Boltzmann distribution, where highly rotational states are excessively populated.
According to the preceding works which report the direct observations of the $X$-state population~\cite{pealat_rovibrational_1985,stutzin__1990,vankan_high_2004,gabriel_formation_2008}, the distribution may have the following properties:
\begin{itemize}
  \item The distribution of the population in the state with small rotational quantum number $N$ in each vibrational state follows the Boltzmann distribution with temperature $T_{rot}^{lo}$.
  \item The large $N$ states are populated more than the Boltzmann distribution and can be approximated by another Boltzmann distribution with higher temperature $T_{rot}^{hi} > T_{rot}^{lo}$.
  \item The total population for each $v$ state follows yet another Boltzmann distribution with $T_{vib}$.
\end{itemize}
Based on these observations, we assume the following two-temperature form for the $X$-state population, 
\begin{align}
  n_{X v N} = n_X \biggl(
    (1 - \alpha) f(E_{X v N} | T_{vib}^{lo}, T_{rot}^{lo})
  +  \alpha f(E_{X v N} | T_{vib}^{hi}, T_{rot}^{hi})
  \biggr),
  \label{eq:Xdistribution}
\end{align}
where $f(E_{X v N} | T_{vib}, T_{rot})$ is the Boltzmann distribution with vibrational and rotational temperatures, $T_{vib}$ and $T_{rot}$, respectively,
\begin{align}
  f(E_{X v N} | T_{vib}, T_{rot}) = 
  \frac{1}{T_{vib}T_{rot}}\exp\left[
      - \frac{E_{X v 0} - E_{X 0 0}}{T_{vib}}
      - \frac{E_{X v N} - E_{X v 0}}{T_{rot}}
    \right].
  \label{eq:two_temperature}
\end{align}
Here, $E_{X v N}$ is the excited energy of $X$ state with vibrational and rotational quantum nubers $v$ and $N$.
Thus, $E_{X 0 0}$ is the ground state energy.
\Eref{eq:Xdistribution} and its substitution to \eref{eq:coronal_model} have six adjustable parameters, i.e., the excitation flux $n_{X}\,R\,n_e$, the weight of the two distributions $\alpha$, and a pair of vibrational and rotational temperatures $T_{rot}^{lo}, T_{vib}^{lo}, T_{rot}^{hi}, T_{vib}^{hi}$.
Here, $n_{X}\,R\,n_e$ represents the product of the total hydrogen density $n_X = \sum_{vN}n_{XvN}$, the electronic-state-resolved excitation rate coefficient $R(R_e)$, and the electron density. 
Because these three terms are unknown and impossible to resolve from the analysis, we treat it as a single adjustable parameter.
We fit the distribution of $n_{d v' N'}$ by adjusting these parameters.
In order to make a robust inference not only for these two frames but also for more than $10^3$ frames of data, we adopt a hierarchical Bayesian model, the detail of which will be described in the Appendix.

The bold curves in \fref{fig:population} show the fit result of $n_{d v' N'}$ for $t$ = 4.00 and 4.75 s.
The change in the populations is well captured.
The reconstructed population in the $X$ state is also shown in the inset.
The estimated values of these parameters are listed in \tref{tb:parameters}.

\begin{table}
  \caption{\label{tb:parameters}Estimated parameters for \#142897 at $t$ = 4.00 and 4.75 s.
  The $\pm$ range indicates the 16--84\% confidential interval of these parameters.
  }
  \begin{tabular}{l c c c c c}
  $t$ & 
  $T_{rot}^{lo}$ (eV) & $T_{rot}^{hi}$ (eV) & 
  $T_{vib}^{lo}$ (eV) & $T_{vib}^{hi}$ (eV) & 
  $\alpha$ 
  \vspace{2mm}
  \\
  \hline
  \vspace{2mm}
  4.00 s & 
  $0.153^{+0.006}_{-0.006}$ & $0.56^{+0.05}_{-0.05}$ & 
  $0.65^{+0.03}_{-0.05}$ & $0.88^{+0.05}_{-0.04}$ &
  $0.48^{+0.02}_{-0.02}$\\
  4.75 s &
  $0.289^{+0.014}_{-0.013}$ & $0.77^{+0.04}_{-0.09}$ &
  $0.77^{+0.04}_{-0.04}$ & $0.90^{+0.04}_{-0.04}$ &
  $0.51^{+0.03}_{-0.03}$ \\
  \end{tabular}
\end{table}

The value of $T_{rot}^{lo}$ changes significantly in these two timings, 
while the changes in $T_{rot}^{hi}, T_{vib}^{lo}, T_{vib}^{hi}$, and $\alpha$ are less significant.
The values of the three temperatures, $T_{rot}^{hi}, T_{vib}^{lo}, T_{vib}^{hi}$, are estimated to the similar range.
As shown in \fref{fig:population} inset, this results in the convergence into a single Boltzmann's distribution for the $X$ state in the highly excited levels.
This behavior is consistent with the previous works, where the population in the $X$ state has been directly measured~\cite{pealat_rovibrational_1985,stutzin__1990,vankan_high_2004,gabriel_formation_2008}.

\section{Parameter dependence of the $X$ state population\label{sec:correlation}}

\begin{figure*}[ht]
  \includegraphics[width=18cm]{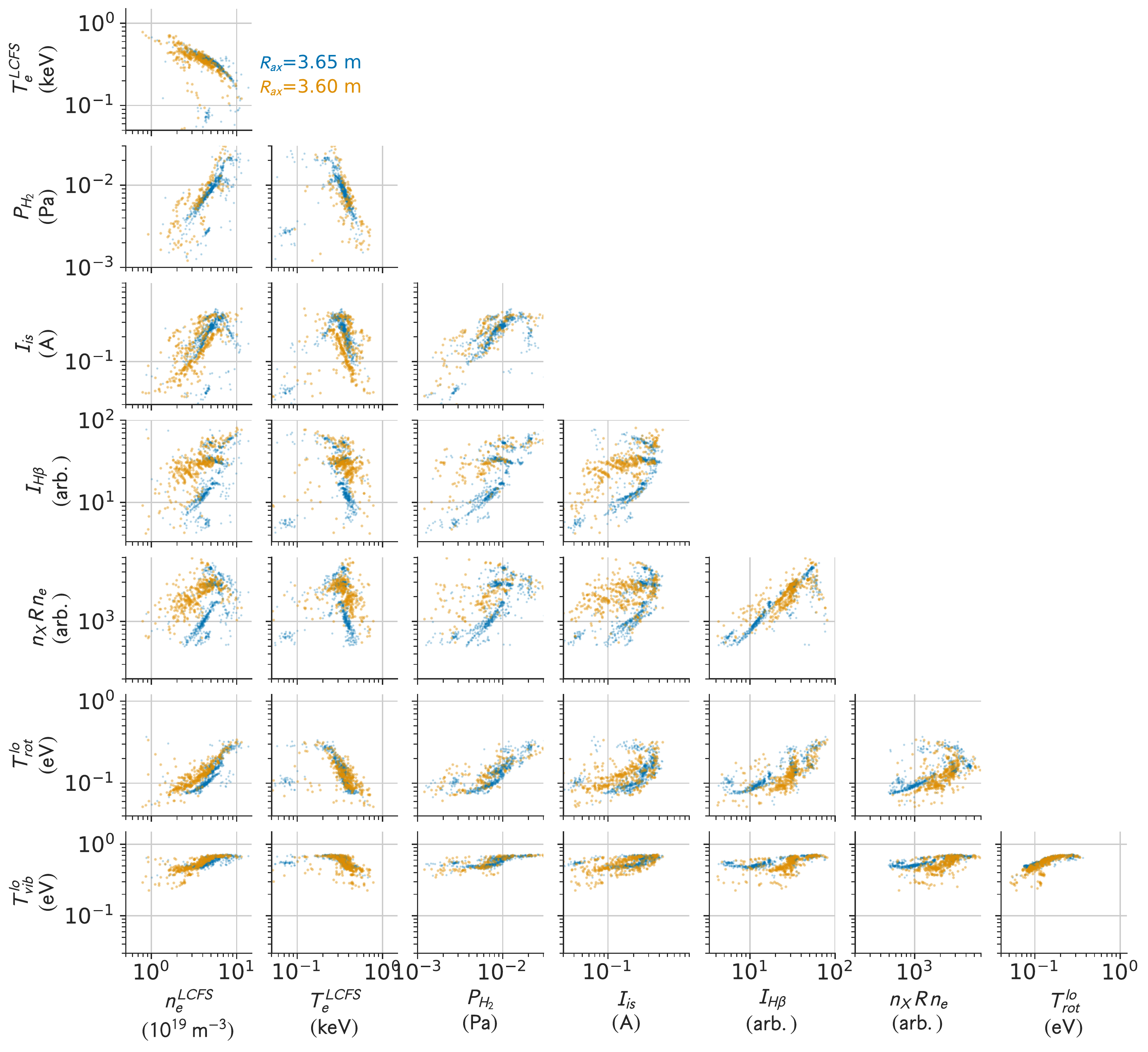}
  \caption{%
  A pair plot for plasma parameters and $T_{rot}^{lo}$ and $T_{vib}^{lo}$. 
  The observation results for different magnetic configurations ($R_{ax}$ = 3.60 m and 3.65 m) are shown in different colors.
  }
  \label{fig:pairplot}
\end{figure*}

We collect the spectra obtained for 120 discharge experiments (74 experiments with $R_{ax}$ = 3.65 m and 46 experiments with $R_{ax}$ = 3.60 m) totalling 1145 frames.
The summary of the experiments is shown in \tref{tb:shotnumber}.
There are also some variations in ECH and NBI powers.
%
The same inference of the $X$ state population is carried out for all of these frames.
\Fref{fig:pairplot} shows the distribution among several parameters, $n_e^{LCFS}$, $T_e^{LCFS}$, $P_{H_2}$, $I_{is}$, $I_{H\beta}$, $n_X\,R\,n_e$, $T_{rot}^{lo}$, and $T_{vib}^{lo}$ at these frames.
The parameters for the $R_{ax} = 3.65$-m- and 3.60-m-experiments are shown in different colors.

\begin{table}[hbt]
  \centering
  \caption{%
    Summary of the experiments we analyze in this work.  
    The ECH column indicates the typical ECH power for these experiments.
    In the column of NBI, the indices of the neutral beam injectors used for the experiments are shown.
    \#1, 2 and 3 are negative-ion based NBI with 180 keV injection energy, which mainly heat the electrons.
    \#4, 5 are the positive-ion based NBI with 40 keV injection energy, mainly heating the ions.
    }
  \label{tb:shotnumber}
  \begin{tabular}{ccccc}
  \hline
  \textit{$R_{ax}$} (m) & shotnumber    & \textit{$B_{t}$} (T) & ECH (MW) & NBI\\
  \hline
  3.65 & 142857-142917 & -2.712& 75& \#1,2,3\\
       & 143523-143560 & 2.712& 60& \#1,2,3\\ 
  \hline
  3.60 & 143293-143297 & -1.640& 0& \#1,2,3,4,5\\
       & 143306-143307 & -2.750& 60&\#1,2,3,4,5\\
       & 143387-143415 & -2.750& 60& \#1,2,3\\
       & 143943-143950 & -2.750& 250-550& \#2,3\\
       & 143973-143988 & -2.750& 75& \#2,3\\
  \hline
  \end{tabular}
\end{table}

\subsection{Plasma parameters}

$n_e^{LCFS}$ varies over 0.06--$1\times 10^{20}\mathrm{\;m^{-3}}$, while $T_e^{LCFS}$ varies 0.08--0.9 keV.
$T_e^{LCFS}$ has a negative correlation against $n_e^{LCFS}$.
The values of $n_e^{LCFS}$ and $T_e^{LCFS}$ are dependent on each other, i.e., the most of the points in the top-left panel in \fref{fig:pairplot} stay on a single curve.
Positive correlations among $n_e^{LCFS}$, $P_{H_2}$, and $I_{is}$ are apparent. 
This may be intuitive, as higher gas pressure often gives more dense plasma and dense plasmas often result in more ion flux to the divertor.
However, the scatter plots show their diffuse relations compared with the $n_e^{LCFS}$-$T_e^{LCFS}$ relation.
This suggests that other factor also affects their relations.
The similar trend can be seen in \fref{fig:shotsummary}~(a), (b), where $n_e^{LCFS}$ and $T_e^{LCFS}$ changes accordingly, while the change in the $P_{H_2}$ and $I_{is}$ are not monotonic against the change of $n_e^{LCFS}$.

The values of $I_{H\beta}$ and $n_X\,R\,n_e$ also show the positive correlations against $n_e^{LCFS}$. 
This is consistent with the positive correlation between $P_{H_2}$ and $n_e^{LCFS}$, i.e., the atom and molecule densities should have a positive dependence on $P_{H_2}$ and the emission rate is almost linearly proportional to $n_e$.
Their relations show a large scatter, suggesting existence of another process to affect the molecular density in front of the divertor.

\subsection{The rotational temperature}

\begin{figure*}[ht]
  \includegraphics[width=18cm]{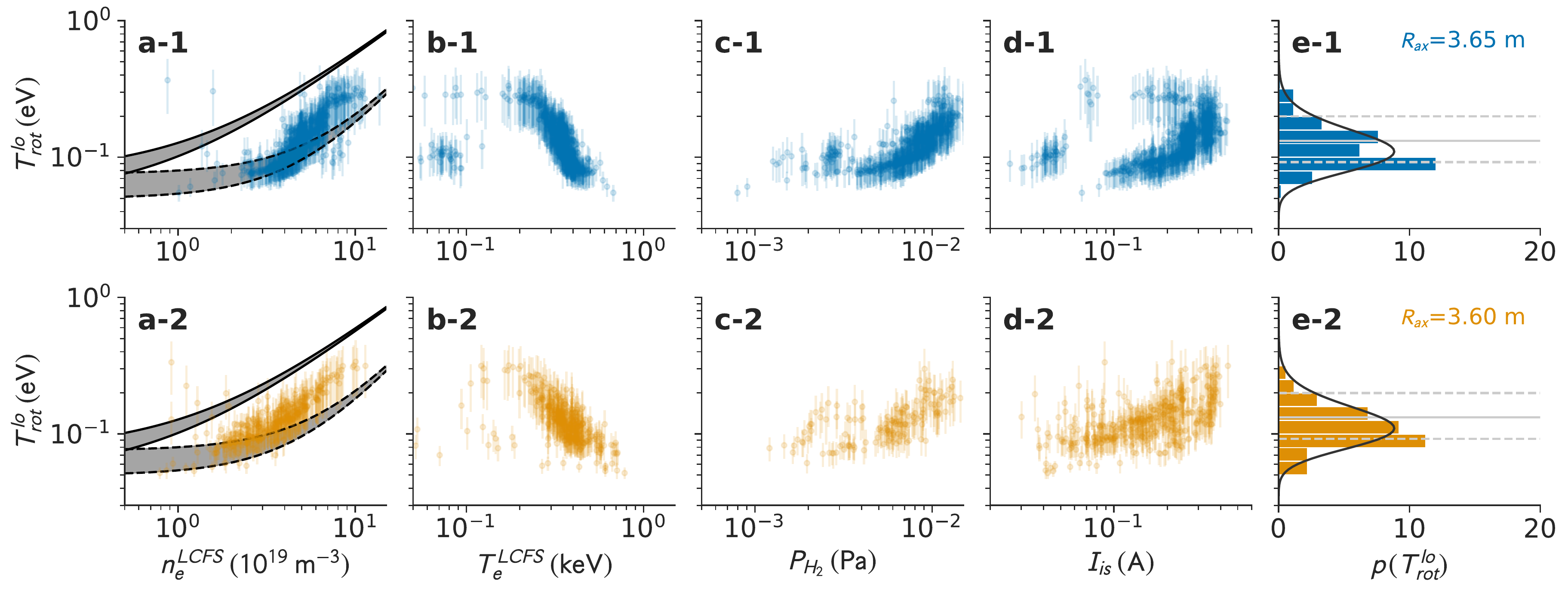}
  \caption{%
  (a-d) Dependence of $T_{rot}^{lo}$ on some plasma parameters.
  The results for $R_{ax}$ = 3.65 m and 3.6 m are shown in the upper and lower panels, respectively.
  Scatters of $T_{rot}^{lo}$-$n_e^{LCFS}$ and $T_{rot}^{lo}$-$T_e^{LCFS}$ are smaller than the rest of plots.
  In (a), the empirical relation of $T_{rot}$ with $n_e$ (\eref{eq:empirical}) is shown by a solid curve.
  The width of the curve indicates the uncertainty in the wall temperature ($300$--$600$ K) at the desorption location.
  The dashed curve is the same relation, but $n_e$ at the dievertor plate is used instead of $n_e^{LCFS}$, which is estimated from \eref{eq:ne_div}.
  (e) The histogram of $T_{rot}^{lo}$. 
  The prior distribution optimized by Bayesian inference is also shown by the solid curve. The median and the 68\% range of the prior are shown by solid and dashed lines, respectively.
  See Appendix for the details of Bayesian inference.
  }
  \label{fig:correlation}
\end{figure*}

The values of $T_{rot}^{lo}$ show a clear dependence on $n_e^{LCFS}$ and $T_e^{LCFS}$.
Expanded correlation plots are shown in \fref{fig:correlation}.
The scatter plot between $T_{rot}^{lo}$ and $n_e^{LCFS}$ is closest to a single curve, while the scatter of $T_{rot}^{lo}$-$P_{H_2}$ and $T_{rot}^{lo}$-$I_{is}$ are larger.
This suggests the direct dependence of the rotational temperature on the electron density, rather than the gas pressure and the ion flux to the divertor.
This interpretation is consistent with that of the previous works~\cite{brezinsek_molecular_2002,unterberg_local_2005,Hollmann2006-zy}.
However, it should be noted that we only consider the electron density at $r_{eff}$ = 0.6 m and those exactly at the emission location are unavailable.

The solid curve in \fref{fig:correlation}~(a) shows the empirical dependence of $T_{rot}$~\cite{Sergienko2013-th,Brezinsek2005-dz,Hollmann2006-zy}, 
\begin{align}
  T_{rot} \approx 280 + T_s + 6 \times 10^{-17} n_e \;\;\mathrm{[K]},
  \label{eq:empirical}
\end{align}
where $T_s$ is the surface temperature (assuming no extra excitation mechanism due to the surface recombination) in K, and $n_e$ is in $\mathrm{m^{-3}}$.
We assume the wall temperature at the desorption location as $T_s \approx 300$--$600$ K.
The width of the curve shows this uncertainty.
Although our result consistently shows the positive $n_e$ dependence with convergence to $\approx$ 600 K  at $n_e \rightarrow 0$, a significant discrepancy is found; our results lie at the larger density side of the solid curve.
This discrepancy may be originated from the difference in $n_e^{LCFS}$ and the $n_e$ values at the emission location.

In order to estimate the electron density on the divertor plate $n_e^{div}$, we use the relation 
\begin{align}
  \label{eq:ne_div}
  \frac{n_e^{div}}{10^{19}\,\mathrm{m^{-3}}} \approx 
  0.08 \times \left[\frac{n_e^{LCFS}}{10^{19}\,\mathrm{m^{-3}}}\right]^{1.5},
\end{align}
which has been suggested in Ref.~\cite{Kobayashi2010-iv}.
The dashed curves in \fref{fig:correlation}~(a) show \eref{eq:empirical} but with this relation, the width of which again indicates $T_s \approx 300$--$600$ K.
Our data point is in between the two curves.
This may be understood that the dominant emission location is between the divertor plate and the LCFS. 
This is consistent with the emission location observation based on Zeeman spectroscopy in~\cite{Fujii2013-ze,Fujii2015-ys}.

\subsection{The vibrational temperature}

\begin{figure*}[ht]
  \includegraphics[width=18cm]{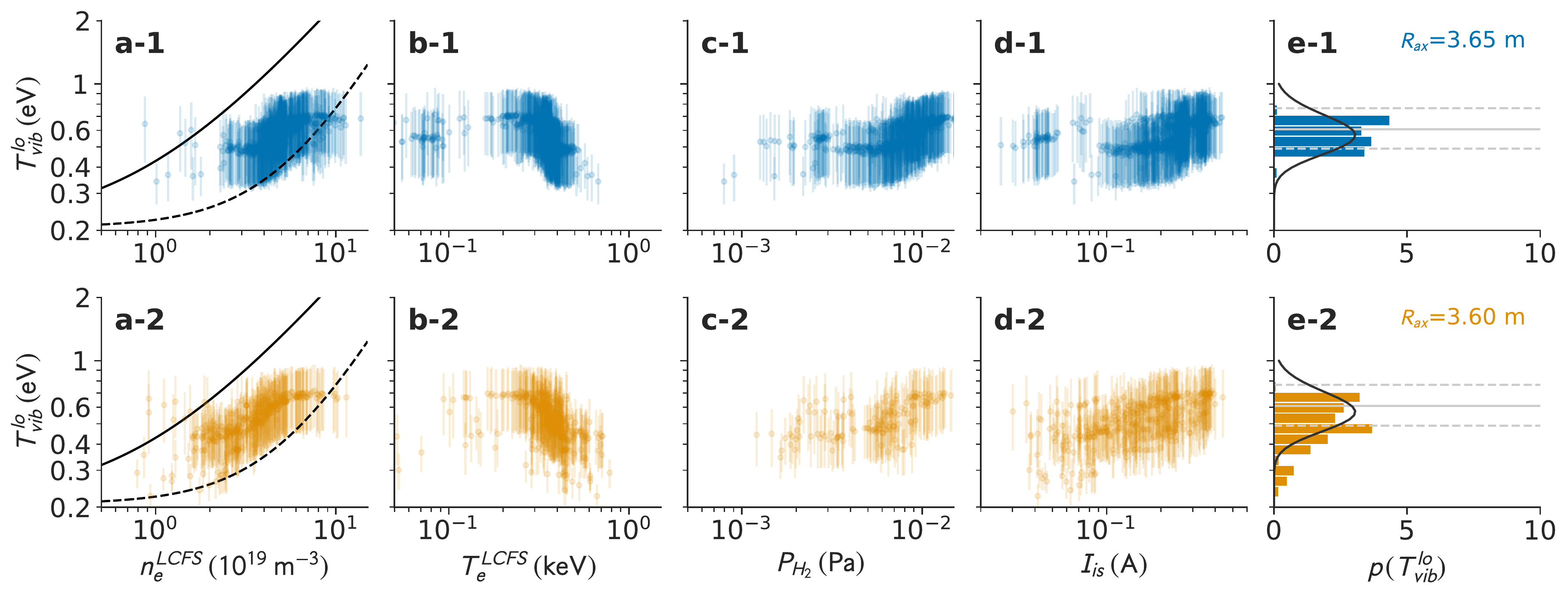}
  \caption{%
  (a-d) Dependence of $T_{vib}^{lo}$ on some plasma parameters.
  The results for $R_{ax}$ = 3.65 m and 3.6 m are shown in the upper and lower panels, respectively.
  In (a), the empirical relation of $T_{vib}$ with $n_e$ (\eref{eq:empirical}) is shown by a solid curve.
  The dashed curve is the same relation, but $n_e$ at the dievertor plate is used instead of $n_e^{LCFS}$, which is estimated from \eref{eq:ne_div}.
  (e) The histogram of $T_{vib}^{lo}$. 
  The prior distribution optimized by Bayesian inference is also shown by the solid curve. The median and the 68\% range of the prior are shown by a solid and dashed lines, respectively.
  See Appendix for the details of Bayesian inference.
  }
  \label{fig:correlation_Tvib}
\end{figure*}

$T_{vib}^{lo}$ also shows a dependence on $n_e^{LCFS}$ and $T_e^{LCFS}$.
Expanded correlation plots are shown in \fref{fig:correlation_Tvib}.
A similar positive $n_e$-dependence of $T_{vib}^{lo}$ has been reported in \cite{brezinsek_molecular_2002}.
From their plot, we extract the dependence as
\begin{align}
  T_{vib} \approx 2400 + 2.6 \times 10^{-16} \; n_e \;\;\mathrm{[K]},
  \label{eq:empirical_Tvib}
\end{align}
with $n_e$ is again in $\mathrm{m^{-3}}$.
The solid curves in \fref{fig:correlation_Tvib}~(a) represent this relation, and the dashed curves show the same relation with the assumption of \eref{eq:ne_div}.
The dependence of $T_{vib}^{lo}$ is in between the two curves.
This is consistent with the above discussion for $T_{rot}^{lo}$, where the emission location of the molecules is suggested between the divertor plate and the LCFS.

These observations for $T_{rot}^{lo}$ and $T_{vib}^{lo}$ indicate that \eref{eq:empirical} and \eref{eq:empirical_Tvib} hold universally, although the geometry of the original experiment is different from that in this work; Brezinsek et al have measured the emission in front of the graphite limiter of TEXTOR~\cite{brezinsek_molecular_2002}, while we measured the emission from LHD divertor.
This suggests that these parameters are mostly determined by the electron density, and the effect of the wall and electron / ion temperatures is small in this parameter range.

\subsection{The other population parameters}

\begin{figure*}[h]
  \includegraphics[width=14cm]{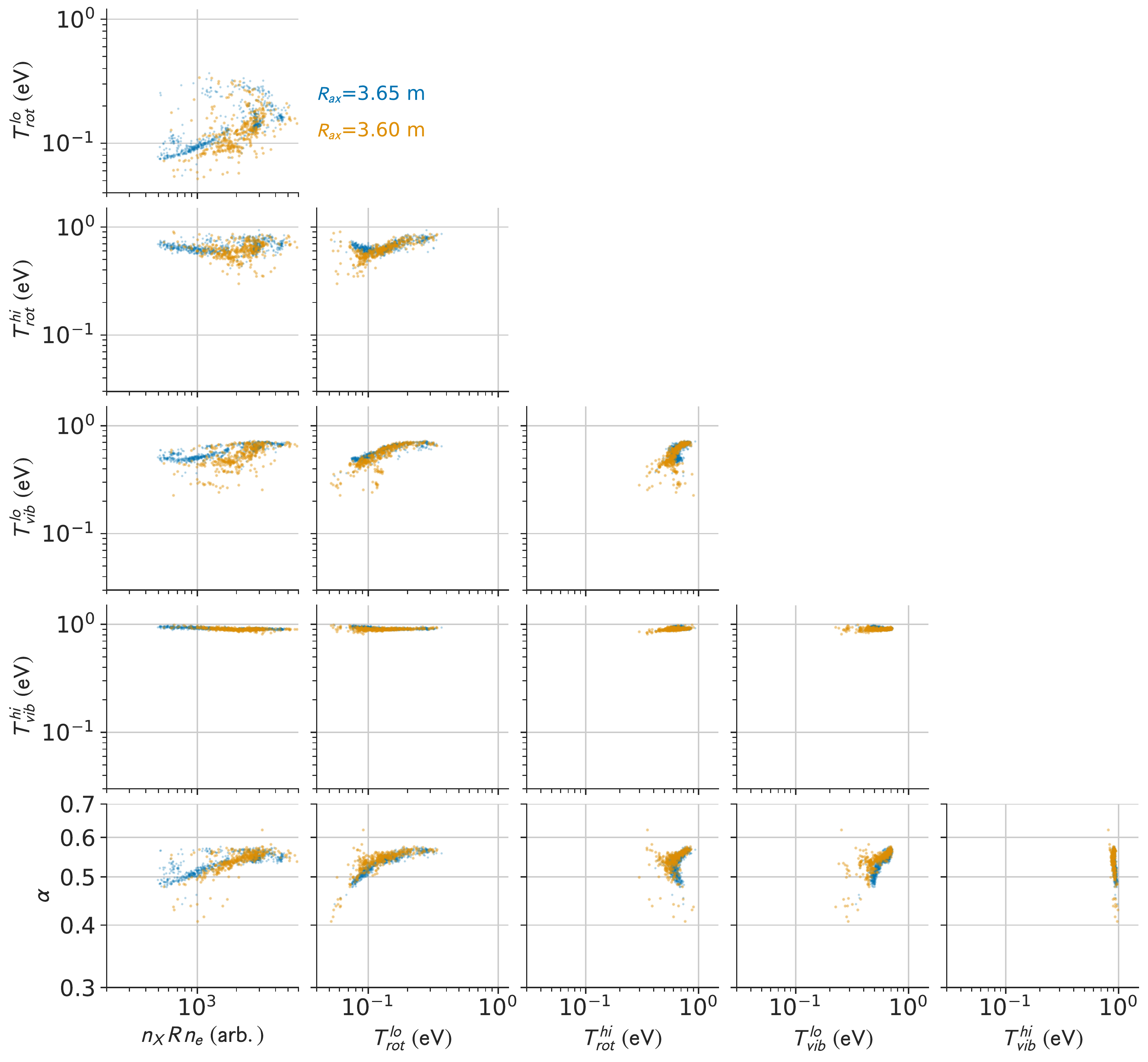}
  \caption{%
  A pair plot for the population parameters.
  The observation results for different magnetic configurations ($R_{ax}$ = 3.60 m and 3.65 m) are shown in different colors.
  The relations among these parameters except for $n_{X}\,R\,n_e$ stay on single curves, suggesting a simple mechanism to determine the population distribution. 
  }
  \label{fig:pairplot2}
\end{figure*}

Correlation among the population parameters, $n_{X}\,R\,n_e$, $T_{rot}^{lo}$, $T_{vib}^{lo}$, $T_{rot}^{hi}$, $T_{vib}^{hi}$, and $\alpha$, are shown in \fref{fig:pairplot2}.
$T_{rot}^{lo}$ and $n_{X}\,R\,n_e$ vary over the factor of 5 and 10, respectively.
On the other hand, the variations of $T_{rot}^{hi}$, $T_{vib}^{lo}$, $T_{vib}^{hi}$, and $\alpha$ are smaller.
For example, $T_{rot}^{hi}$ only changes by a factor of $\approx$ 1.5.
Furthermore, the scatter plots among these population parameters except for $n_{X}\,R\,n_e$ show a significant correlation.
For example, the pair plot of $T_{rot}^{lo}$ and $n_{X}\,R\,n_e$ (the top left panel) shows a large scatter, while $T_{rot}^{lo}$-$T_{vib}^{lo}$ (the second panel from the top and the left) is close to a single curve.
This suggests that a simple process determines the entire population distribution.

There are several processes have been proposed for the $X$ state population of hydrogen molecules, such as electron impact, proton impact, and neutral particle impact~\cite{sawada_rovibrationally_2016}.
From \fref{fig:pairplot}, we see high correlation between $T_{rot}^{lo}$ and $n_e$.
On the other hand, the correlation between $T_{rot}^{lo}$ and $n_{X}\,R\,n_e$ shows larger scatter, although the $T_e$-dependence of $R$ is known to small.
This indicates that the dependence on $n_{X}$ is smaller, suggesting the unimportance of the molecular-collision effect on the population distribution.
As such, a further study based on our data may help identification of important processes.

\section{Conclusion\label{sec:conclusion}}

We analyzed a thousand visible spectra of Fulcher-$\alpha$ band measured for divertor plasmas in the LHD.
With the coronal model and Baysian inference, the population distribution of the electronical ground state was estimated from these spectra.
A nearly linear $n_e$-dependences of $T_{rot}^{lo}$ and $T_{vib}^{lo}$ were found, which is consistent with previous works.
On the other hand, the higher rotational and vibrational temperatures as well as the mixture coefficient only show slight changes over the broad range of plasma parameters, and these parameters show a significant correlation.
This suggests an importance of electron-impact excitation to determine the non-thermal population distribution of molecular hydrogen in divertor plasmas.

\appendix

\section{Bayesian inference of the $X$-state population}

The two-temperature model we assume in \eref{eq:two_temperature} is not always robust.
For example, the two temperature model has many optima that fit equally well for a single-temperature distribution; $T^{lo} = T^{hi}$ with any value of $\alpha$ is one optimum, but $\alpha = 0$ with any value of $T^{hi}$ is another optimum.
In order to carry out a robust inference on the population parameters in \eref{eq:two_temperature} for thousands of frames, we utilized Bayesian inference framework.
In this section, we show the details of the probabilistic model we have adopted in this work.

Let $\mathbf{n}=\{n_{t,l} | f\in \{1, 2, \cdots, N_t\}, l\in \{1, 2, \cdots, N_l\}\}$ be the observed populations at the $d$ state and $\sigma_{t,l}$ be its uncertainty from the measurement, where $N_t$ is the total numbef of frames and $N_l$ is the total number of molecular lines.for these spectra
For each frame $t$, we define the population parameters of the $X$ state, $\theta_t = \{[n_X R n_e]_t, T_{rot}^{lo} {_t}, T_{vib}^{lo} {_t}, T_{rot}^{hi} {_t}, T_{vib}^{hi} {_t}, \alpha_t\}$.
Our aim is to infer the posterior distribution of $\theta_t$ with given the observation $\mathbf{n}$, $p(\theta_t | \mathbf{n})$.

With a given value of $\theta_t$, we can compute the expected populations $\overline{n}_{l}(\theta_t)$ in the $d$ state based on \eref{eq:coronal_model}.
We assume the gaussian noise for $n_{t,l}$ with the standard deviation $\gamma\sigma_{t,l}$,
\begin{align}
  p(n_{t, l} | \theta_t) = \mathcal{N}(n_{t, l} | \overline{n}_{l}(\theta_t), \gamma \sigma_{t, l})
\end{align}
where $\mathcal{N}(x | \mu, \sigma) = \exp[-(x-\mu)^2/2\sigma^2] / \sqrt{2\pi}\sigma$ is the normal distribution with mean $\mu$ and the standard deviation $\sigma$.
Here, we additionally assume the excess noise factore $\gamma$, which is a single parameter common for all the frames and upper states to capture the possible unidentified error in the measurement, e.g. overlap by other lines.
$\gamma$ is estimated and merginalized later.

\begin{figure}[bt]
  \includegraphics[width=8cm]{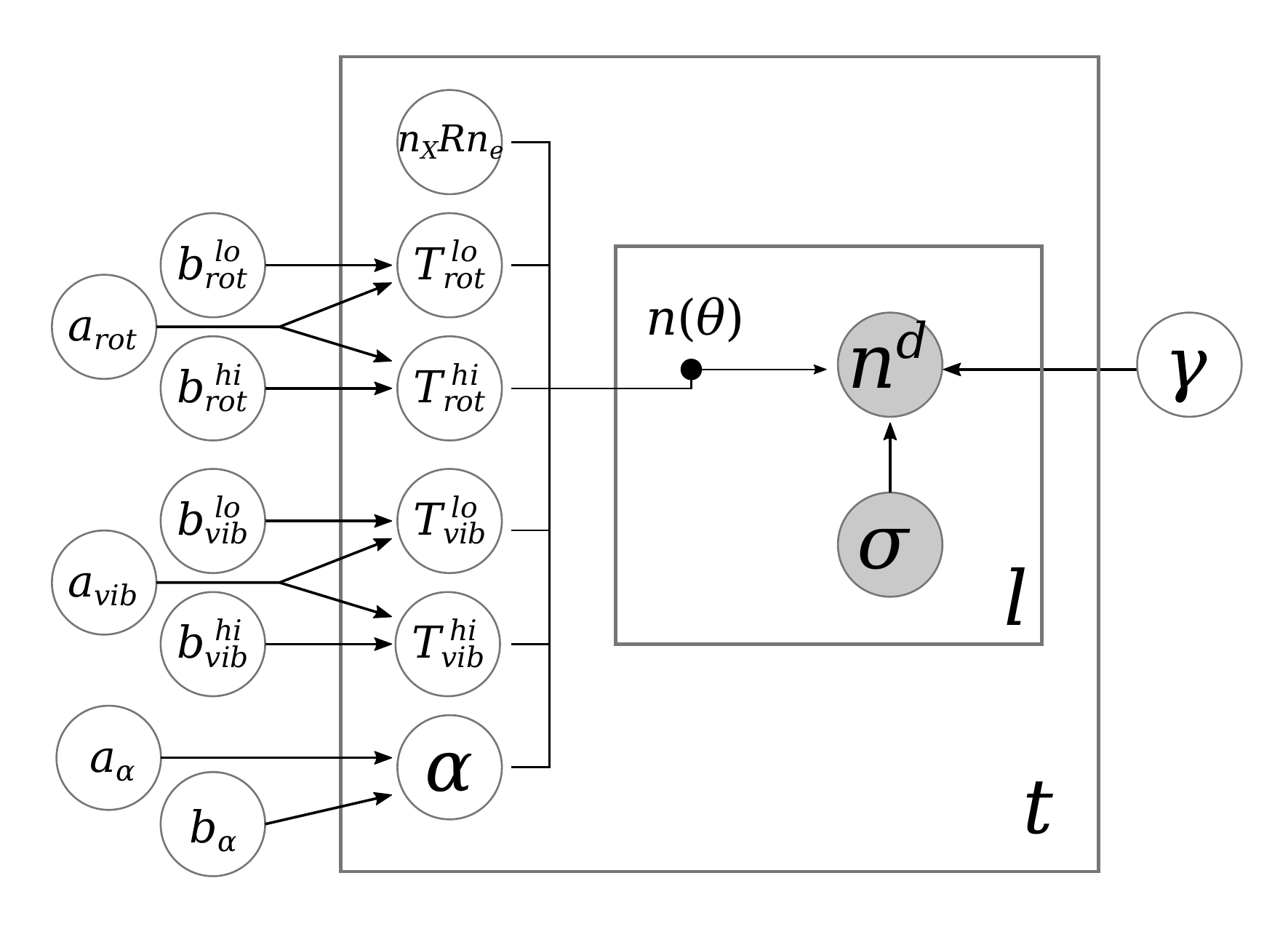}
  \caption{%
  A graphical representation of our hierarchical Bayesian model.
  The measured quantities are indicated by filled circles, while the inferred random variables are shown by open circles.
  A variable shown with a dot ($n(\theta)$) represents a fixed parameter with given input parameters.
  Square panels indicate that variables inside a panel have vector-valued parameters.
  For example, population parameters, such as $T_{rot}^{lo}$, are shared by all the upper states $l$, while hyperparameters such as $a_{rot}$ are shared for all the frames $t$.
  }
  \label{fig:graphical}
\end{figure}

We also assume prior distributions for the population parameters $\theta_t$, which makes the inference robust.
As a prior for $[n_X R n_e]_t, T_{rot}^{lo} {_t}, T_{vib}^{lo} {_t}, T_{rot}^{hi} {_t}$, and $T_{vib}^{hi} {_t}$, we assume the inverse gamma distribution, which is a typical distribution for a nonnegative random variable,
\begin{align}
  p([n_X R n_e]_t | a_n, b_n) &= \mathcal{IG}([n_X R n_e]_t | a_n, b_n)\\
  p(T_{rot}^{lo} {_t} | a_{rot}, b_{rot}^{lo}) &= \mathcal{IG}(T_{rot}^{lo} {_t} | a_{rot}, b_{rot}^{lo})\\
  p(T_{rot}^{hi} {_t} | a_{rot}, b_{rot}^{hi}) &= \mathcal{IG}(T_{rot}^{hi} {_t} | a_{rot}, b_{rot}^{hi})\\
  p(T_{vib}^{lo} {_t} | a_{vib}, b_{vib}^{lo}) &= \mathcal{IG}(T_{vib}^{lo} {_t} | a_{vib}, b_{vib}^{lo})\\
  p(T_{vib}^{hi} {_t} | a_{vib}, b_{vib}^{hi}) &= \mathcal{IG}(T_{vib}^{hi} {_t} | a_{vib}, b_{vib}^{hi}),
\end{align}
and for $\alpha_t$ we assume beta distribution, which is a typical distribution for a variable in $[0,1]$,
\begin{align}
  p(\alpha_t | a_n, b_n) &= \mathcal{B}(\alpha_t | a_\alpha, b_\alpha)
\end{align}
where $\mathcal{IG}(x|a,b) = x^{-a -1}e^{-b / x}\,b ^{a} / \Gamma(a)$ is the inverse gamma distribution with the shape parameter $a$ and the scale parameter $b$ with $\Gamma(a) = \int_0^\infty x^{a-1}e^{-x}dx$ the gamma function.
$\mathcal{B}(x|a,b) = x^{a-1} (1-x)^{b-1} / B(a,b)$ with the beta function $B(a,b) = \int_0^1 x^{a-1} (1-x)^{b-1} dx$.
The hyperparameters, $\Theta = \{\gamma, a_n$, $b_n$, $a_{rot}$, $b_{rot}^{lo}$, $b_{rot}^{hi}$, $a_{vib}$, $b_{vib}^{lo}$, $b_{vib}^{hi}$, $a_\alpha$, $b_\alpha\}$ are common for all the frames and will be inferred simultaneously.
The graphical representation for our model is shown in \fref{fig:graphical}.

Based on the Bayes rule, the posterior distribution can be found by integrating the hyperparameters,
\begin{align}
  p(\theta_t | \mathbf{n}, \boldsymbol{\sigma}) \propto 
  \int d\Theta
  \prod_{t,l} p(n_{t,l} | \overline{n_t}(\theta_t), \gamma \sigma_{t,l})
  p(\theta_t | \Theta)p(\Theta),
\end{align}
where $p(\Theta) = \Gamma(\Theta | 1, 1)$ is the hyperprior distribution for $\Theta$.
The inference (i.e., the integration) was carried out with Markov Chain Monte Carlo (MCMC) method with \texttt{Stan Modeling Language}~\cite{carpenter2017stan,standev2022pystan}.

The median values of the hyperparameters are shown in \tref{tb:hyperpriors}.
The prior distributions of $T_{rot}^{lo}$ and $T_{vib}^{lo}$ are shown in \fref{fig:correlation}~(e) and \fref{fig:correlation_Tvib}~(e), respectively, by solid curves. 
The histogram of the estimated $T_{rot}^{lo}$ and $T_{vib}^{lo}$ are also shown in the same panel.
It can be seen that these prior distributions are optimized so that they fit the entire distributions of $T_{rot}^{lo}$ and $T_{vib}^{lo}$.
In this panel, the median value as well as the $68$\% intervals are also shown.

\begin{table}
  \caption{\label{tb:hyperpriors}Inferred values of the hyperprior parameters for our Bayesian model.
  The range for the 68\% confidential interval for the corresponding parameters are also tabulated.
  }
  \begin{tabular}{l c c c}
  parameter & $a$ & $b$ & 16\% -- 50\% -- 84\% \\
  \hline
  $T_{rot}^{lo}$ (eV) & 7.0 & 0.88 & 0.09 -- 0.13 -- 0.20
  \\
  $T_{rot}^{hi}$ (eV) &     & 4.5  & 0.47  -- 0.67 -- 1.01
  \\
  $T_{vib}^{lo}$ (eV) & 20 & 12 & 0.49 -- 0.61 -- 0.77
  \\
  $T_{vib}^{hi}$ (eV) &     & 18 & 0.74 -- 0.92 -- 1.15
  \vspace{2mm}
  \\
  $n$ (arb.) & 0.85 & 3 & 2 -- 6 -- 30 \\
  $\alpha$ & 36 & 54 & 0.35 -- 0.40 -- 0.45
  \end{tabular}
\end{table}

\section*{acknowledgments}
  This work was supported by JSPS KAKENHI Grant Number 19K14680.
\bibliographystyle{plain}
\bibliography{refs}

    
\end{document}